
\pageno=0
\magnification=1000
\rightline{CERN-TH.6553/92}
\rightline{ILL-(TH)-92-13}
\baselineskip =20pt
\vskip1.5truecm
\centerline
{\bf THE FOUR-FERMI MODEL IN THREE DIMENSIONS}
\centerline
{\bf AT NON-ZERO DENSITY AND TEMPERATURE }
\vskip1truecm

\centerline{Simon HANDS}
\vskip 0.3 truecm
\centerline{\it Department of Physics and Astronomy, University of Glasgow,
Glasgow G12 8QQ, U.K.}

\centerline{\it and}

\centerline{{\it Theory Division, CERN,
CH-1211 Geneva 23, Switzerland}\footnote{$\,^{ }$}{CERN-TH.6553/92}}
\vskip0.5truecm

\centerline{Aleksandar KOCI\'C}
\vskip 0.3 truecm
\centerline{{\it Theory Division, CERN,
CH-1211 Geneva 23, Switzerland}\footnote{$\,^{ }$}{ILL-(TH)-92-13}}
\vskip 0.5 truecm

\centerline{John B. KOGUT}
\vskip .3 truecm
\centerline{\it Department of Physics, University of Illinois at
Urbana-Champaign}
\centerline{\it 1110 West Green Street, Urbana, IL 61801, U.S.A.}
\vskip 1.0 truecm
\centerline{\bf Abstract}

{\narrower
The Four Fermi model with discrete chiral symmetry is studied in three
dimensions at non-zero chemical potential and temperature using the Hybrid
Monte Carlo algorithm.  The number of fermion flavors is chosen large
$(N_f =
12)$ to compare with analytic results.  A first order chiral symmetry
restoring transition is found at zero temperature with a critical chemical
potential $\mu_c$ in good agreement with the large $N_f$ calculations.  The
critical index $\nu$ of the correlation length is measured in good agreement
with analytic calculations.  The two dimensional phase diagram (chemical
potential vs. temperature) is mapped out quantitatively.  Finite size
effects on relatively small lattices and non-zero fermion mass effects are
seen to smooth out the chiral transition
dramatically.\footnote{$\,^{ }$}{June 1992}
\smallskip}
\vfill\eject

\noindent{\bf 1. Introduction}

The behavior of symmetries at finite temperatures and densities is one of
the most outstanding and relevant problems in many current areas of
particle physics;
e.g. cosmology, relativistic heavy-ion collisions, and the quark-gluon
plasma [1]. In recent years we have witnessed revived interest in the chiral
symmetry restoration transition in QCD. The problem of symmetry breaking and
its
restoration is intrinsically nonperturbative. Therefore, the number of
available
techniques is limited and most of our knowledge about the phenomenon comes from
lattice simulations. Due to the enormous complexity of QCD, studies have so
far been done on lattices of modest size  and have been unable to
yield quantitative claims as far as the order of the transition is concerned.
This is unfortunate since several studies suggest that the high temperature
phase of QCD has a number of interesting features [2]. In addition, only very
slow progress has been made in lattice simulations at finite chemical
potential,
which is the regime of direct relevance to real physics [3].

In this paper we will approach the general problem of chiral symmetry
restoration
at finite temperature and density in a three-dimensional toy model in order
to understand what
ingredients might play a decisive role in more complex systems like gauge
theories. We have simplified our model as much as possible
in order to produce the highest quality data and learn what range of parameters
we need for studies of more realistic cases.
We have thus chosen to study the Gross-Neveu model [4], in which the chiral
symmetry is discrete. The Lagrangian is
$${\cal L}=\sum_{j=1}^{N_f}\left[\bar\psi^{(j)}\partial{\!\!\! /}\,\psi^{(j)}
-{g^2\over{2N_f}}(\bar\psi^{(j)}\psi^{(j)})^2\right].\eqno(1.1)$$
Here $\psi^{(j)}$ is a four component spinor, $N_f$ is the number of flavors
of elementary fermion, and the discrete ${\rm Z_2}$
chiral symmetry is $\psi\mapsto\gamma_5\psi$,
$\bar\psi\mapsto-\bar\psi\gamma_5$. Our choice of model also reflects our
interest in its behavior at zero temperature and density [5,6]: in less than
four dimensions the model has a non-trivial renormalization group fixed
point, also characterized by the spontaneous breaking of the chiral symmetry
at a critical coupling $g^2_*$.
It is thus a toy model for the study of non-trivial strongly coupled theories.
Although the theory is non-renormalizable in a standard perturbation expansion
in the coupling $g^2$, its $1/N_f$ expansion about the fixed point $g^2_*$
is renormalizable. We have argued elsewhere [6] that this is closely related
to the fact that the theory's critical indices satisfy hyperscaling. Physically
this means that the theory has a divergent correlation length at the fixed
point and this length sets the scale for the theory's low-energy phenomena.
To $O(1/N_f)$ the theory's critical indices are [6]
$$\nu=1+{8\over{3N_f\pi^2}};\;\;\delta=2+{8\over{N_f\pi^2}};\;\;
\beta_m=1;\;\;\gamma=1+{8\over{N_f\pi^2}};\;\;\eta=1-{16\over{3N_f\pi^2}},
\eqno(1.2)$$
in the standard notation of classical statistical mechanics [7].

There are several motivations for studying such a simple model.
In four dimensions the four-fermi model is believed to be an effective theory
of quarks and gluons at intermediate energies. The degrees of freedom are
light mesons and quarks. As far as finite temperature and density is concerned,
the low temperature regime will be dominated by the lightest particles and, if
the restoration temperature is of the order of 100MeV, then the
contribution of the heavier particles like $\rho$ mesons will be exponentially
suppressed. In that sense, the universal properties of chiral symmetry
restoration in QCD could be
well described by an effective theory like the Nambu -- Jona-Lasinio model.
In three dimensions, however, the four-fermi model is actually renormalizable,
and precise analytic predictions are available from an expansion
in $1/N_f$. Although we consider the simplest such model in this paper,
extensions to more realistic models where the chiral symmetry is continuous are
straightforward [4].
One might think of a yet more drastic simplification, and consider the model
in two dimensions. However, in this case there are conceptual problems; eg.
in the ${\rm Z}_2$ case the symmetry restoration is now dominated by
the materialization of kink -- anti-kink states, which are composite states of
the fundamental fermion fields, which are not probed in the $1/N_f$ expansion.
In two dimensions the extension to continuous symmetries is also plagued by
the non-existence of massless Goldstone bosons [8].

The major technical barrier to progress in
simulating non-vanishing baryon number densities
in QCD is the absence of a probabilistic interpretation of the path integral
measure due to the action becoming complex once the chemical potential
$\mu\not=0$. In the model considered here, the
action remains real even after the introduction of $\mu$, which means
we can study the physics of the high-density regime using standard Monte Carlo
techniques. In addition, it has fewer degrees of freedom than QCD, or even the
Nambu -- Jona-Lasinio model,
and hence can be studied with greater precision on much bigger lattices than
are presently used for QCD thermodynamics.
In general, simulations focus on the
evaluation of the order parameter $<\bar\psi\psi>$ which requires the inversion
of the Dirac operator. Most numerical algorithms which simulate the effect
of virtual quark -- anti-quark pairs in the vacuum also require the inversion
of this operator; this is the most computer-intensive step in such simulations.
In gauge theories, this operator is singular in the
chiral limit (ie. the matrix to be inverted becomes ill-conditioned),
and the simulations have to be done using a finite bare mass. Information
about the chiral limit is then obtained by extrapolating the finite mass data
to $m=0$. This last step is severely constrained by the lattice volume;
$m$ can not be taken arbitrarily small, since otherwise the Compton
wavelength of the Goldstone pion associated with the symmetry breaking would
exceed the lattice size producing severe finite volume effects [9].
Consequently, lattice QCD
data, always taken at finite mass, show only a crossover rather
than a real phase transition. In our model, we are dealing with a Yukawa-like
coupling (see below),
so that the fermion matrix to be inverted has non-zero diagonal
elements even when the bare mass is set to zero. The order
parameter (the inverse Dirac operator) is thus not singular
and the simulations can be done in the chiral limit directly. In this way, we
can in principle
explore the systematics of finite mass effects, finite volume effects,
and the validity of various extrapolation procedures. This
might teach us how to use the QCD data better.

The mean field theory
description of the transition, to be presented in Sec. 3, predicts a first
order transition for $T=0$ and a continuous transition for $T>0$.
Both analytic and numerical
work in this model is aided by the introduction of an auxiliary scalar field
$\sigma$, so eqn. (1.1) becomes
$${\cal L}=\sum_{j=1}^{N_f}\left[\bar\psi^{(j)}\partial{\!\!\! /}\,\psi^{(j)}
+\sigma\bar\psi^{(j)}\psi^{(j)}\right]+{N_f\over{2g^2}}\sigma^2,\eqno(1.3)$$
and the Lagrangian becomes quadratic in the fermion field. The ground state
expectation value of $\sigma$ serves as a convenient order parameter for the
theory's critical point. Using the standard lattice regularization scheme to
be discussed in Sec.2, the bulk critical coupling is found to be
$1/g^2_*=0.975$ - 1.000 for $N_f=12$ when the chemical potential is set to
zero as in eqn. (1.3) [5]. Simulation of $20^3$ lattices at various couplings
close to the critical point (within its scaling region as determined in ref.
[5]) showed that $\Sigma=<\sigma>$ jumps {\it discontinuously\/} to zero as
$\mu$ is increased and that the induced fermion density also jumps
{\it discontinuously\/} through the transition. In addition, the dependence of
$\mu_c$ on the coupling allows us to determine the correlation length exponent
$\nu$ for $N_f=12$:
$$\nu=1.05(10),\eqno(1.4)$$
in good agreement with the large $N_f$ prediction (1.2). Furthermore the
magnitude of $\mu_c$ itself is found to be in good agreement with the mean
field result $\mu_c=\Sigma_0$, where $\Sigma_0$ is the value of the vacuum
expectation of the scalar field at zero temperature and chemical potential.
This result indicates that materialization of the fermion itself drives the
symmetry restoration transition: this is not the case in the two-dimensional
Gross-Neveu model, where kink -- anti-kink states materialize at the transition
[10,11]. The fact that $\mu_c$ agrees with the mean field result is indirect
evidence that such exotic fermion states in which the energy per constituent is
smaller than the energy of a single fermion state do not occur in the three
dimensional model [12].

We also simulated the model at both non-zero $\mu$ and non-zero temperature
$T$. In ref. [5] we confirmed that the $\mu=0$, $T\not=0$ symmetry restoring
transition occurs at $T_c/\Sigma_0\simeq0.72$ in good agreement with mean field
predictions, and that this transition is second order. Here we map out the
phase diagram in the $(\mu,T)$ plane.

The simplicity of this model and the efficiency of the simulation algorithm
allowed us to address two technical issues of interest to lattice gauge
theorists studying QCD in extreme environments. First, four-dimensional QCD
simulations are presently restricted to relatively small lattices. We show here
that if the lattice is taken relatively small ($12^3$ as opposed to $20^3$)
the non-zero $\mu$ transition is
smeared out and all evidence for a discontinuous
transition is lost. Furthermore, as discussed above, QCD cannot be simulated
directly in the chiral limit. We show here that, even on relatively large
lattices and at couplings chosen in the scaling region, even very small bare
fermion masses $m$ smooth the transition dramatically. We learn that it would
be very difficult indeed to find evidence for a discontinuous transition by
approaching the chiral limit $m=0$ of the theory via $m\not=0$ simulations.

This paper is organized into several sections. In Sec. 2 we briefly discuss the
lattice formulation of the model and its simulation algorithm. More detail on
these issues has already been provided in ref [5]. In Sec. 3 we present the
mean field analysis of the theory -- this overlaps to some extent with the
results of ref. [12]. In Sec. 4 the simulation study of the zero-temperature
model at non-zero chemical potential is presented. In Sec. 5 the model is
sudied at both $\mu\not=0$ and $T\not=0$. Finally, in Sec. 6 we consider finite
volume effects by simulating on a $12^3$ lattice, and finite $m$ effects on
relatively large lattices. In both cases the discontinuous nature of the
transition driven by non-zero chemical potential is lost. Sec. 7 summarizes
our work.
\vskip1truecm

\noindent{\bf2. Lattice Formulation of the Gross-Neveu Model}

The Gross-Neveu model in its bosonised form (1.3) may be formulated
on a space-time lattice using the following action:
$$S=\sum_{i=1}^{N_f/2}\left(\sum_{x,y} \bar\chi_i(x){\cal M}_{x,y}\chi_i(y)
+{1\over8}\sum_x\bar\chi_i(x)\chi_i(x)\sum_{<{\tilde x},x>}\sigma({\tilde x})
\right)+{N_f\over{4g^2}}\sum_{\tilde x}\sigma^2({\tilde x}),\eqno(2.1)$$
where $\chi_i,\bar\chi_i$ are Grassmann-valued staggered fermion fields
defined on the lattice sites, the auxiliary scalar field $\sigma$ is defined
on the dual lattice sites, and the symbol $<{\tilde x},x>$ denotes the set of 8
dual lattice sites ${\tilde x}$ surrounding the direct lattice site $x$.
The lattice spacing $a$ has been set to one for convenience. The
fermion kinetic operator ${\cal M}$ is given by:
$${\cal M}_{x,y}={1\over2}\left[\delta_{y,x+\hat0}e^\mu-
\delta_{y,x-\hat0}e^{-\mu}\right]+{1\over2}\sum_{\nu=1,2}\eta_\nu(x)
\left[\delta_{y,x+\hat\nu}-\delta_{y,x-\hat\nu}\right],\eqno(2.2)$$
where $\eta_\nu(x)$ are the Kawamoto-Smit phases $(-1)^{x_0+\cdots+x_{\nu-1}}$.
The influence of the chemical potential $\mu$ is manifested through the
timelike
links, following [13]: only fermion loops which wrap around the timelike
direction are affected by its inclusion.

The Gross-Neveu model in two dimensions was first formulated using auxiliary
fields on the dual sites in reference [14]. We can motivate this particular
scheme by considering a unitary transformation to fields $u$ and $d$ [15]:
$$\eqalign{
u_i^{\alpha a}(Y)&={1\over{4\surd2}}\sum_A\Gamma_A^{\alpha a}\chi_i(A;Y), \cr
d_i^{\alpha a}(Y)&={1\over{4\surd2}}\sum_A B_A^{\alpha a}\chi_i(A;Y). \cr}
\eqno(2.3)$$
Here $Y$ denotes a site on a lattice of twice the spacing of the original, and
$A$ is a lattice vector with entries either 0 or 1, which
ranges over the corners of the elementary cube associated with $Y$, so that
each site on the original lattice corresponds to a unique choice of $A$ and
$Y$.
The $2\times2$ matrices $\Gamma_A$ and $B_A$ are defined by
$$\eqalign{
\Gamma_A&=\tau_1^{A_1}\tau_2^{A_2}\tau_3^{A_0},\cr
B_A&=(-\tau_1)^{A_1}(-\tau_2)^{A_2}(-\tau_3)^{A_0},\cr}\eqno(2.4)$$
where the $\tau_\nu$ are the Pauli matrices. Now, if we write
$$q_i^{\alpha a}(Y)=\left(\matrix{u_i^\alpha(Y)\cr d_i^\alpha(Y)\cr}\right)^a,
\eqno(2.5)$$
and interpret $q$ as a 4-spinor with two flavors counted by the latin index
$a$,
then the fermion kinetic term of the action (2.1) may be recast in Fourier
space as follows:
$$\eqalign{
S_{kin}=\int{{d^3k}\over{(2\pi)^3}}\sum_i\sum_{\nu=1,2}{i\over2}\bigg\{
&\bar q_i(k)(\gamma_\nu\otimes1_2)q_i(k) \sin2k_\nu+
 \bar q_i(k)(\gamma_5\otimes\tau_\nu^*)q_i(k) (1-\cos2k_\nu)\bigg\}
\cr
+{1\over2}\bigg\{&\bar q_i(k)(\gamma_0\otimes1_2)q_i(k) \big[i\sin2k_0\cosh\mu
            +(1+\cos2k_0)\sinh\mu\big]\cr
+&\bar q_i(k)(\gamma_5\otimes\tau_3^*)q_i(k)\big[i(1-\cos2k_0)\cosh\mu
            +\sin2k_0\sinh\mu\big]\bigg\},\cr}\eqno(2.6)$$
where
$$(\gamma_\nu)_{\alpha\beta}=
\left(\matrix{\tau_\nu&\cr&-\tau_\nu\cr}\right)_{\alpha\beta};\;\;
(\gamma_0)_{\alpha\beta}
=\left(\matrix{\tau_3&\cr&-\tau_3\cr}\right)_{\alpha\beta};
\;\;(\gamma_5)_{\alpha\beta}
=\left(\matrix{&-i1_2\cr i1_2&\cr}\right)_{\alpha\beta},\eqno(2.7)$$
the second set of ($2\times2$) matrices in the direct product act on the flavor
indices, and the momentum integral extends over the range
$k_\nu\in(-\pi/2,\pi/2]$. At non-zero temperature the lattice has finite
extent in the temporal direction, and $\int dk_0$ is replaced by a sum over
$N_\tau /2$ modes, where $N_\tau$  is the number of lattice spacings in the
time direction, and antiperiodic boundary conditions are imposed on the fermion
fields. In the classical continuum limit lattice spacing $a\to0$, the flavor
non-diagonal terms vanish as $O(a)$, and we recover the standard Euclidian
form $\bar q_j(\partial{\!\!\! /}\,+\mu\gamma_0)q_j$, where the flavor index
$j$ now runs from 1 to $N_f$.

Similarly, it is straightforward to show that the interaction terms can be
rewritten (with obvious notation):
$$S_{int}=\sum_Y\left(\sum_A\sigma(A;{\tilde Y})\right)
\bar q_i(Y)(1_4\otimes1_2)q_i(Y)+O(a),\eqno(2.8)$$
where the $O(a)$ terms contain non-covariant and flavor non-singlet terms. If
we used a formulation in which the $\sigma$ fields lived on the direct lattice
sites, then such non-covariant terms would contribute at $O(a^0)$ [14].

Thus we see that the lattice action (2.1) reproduces the bosonised Gross-Neveu
model at non-zero density, at least in the classical continuum limit. Most
importantly, (2.1) has a discrete global invariance under
$$\chi_i(x)\mapsto(-1)^{x_0+x_1+x_2}\chi_i(x);\;\;
\bar\chi_i(x)\mapsto-(-1)^{x_0+x_1+x_2}\bar\chi_i(x);\;\;
\sigma({\tilde x})\mapsto-\sigma({\tilde x}),\eqno(2.9a)$$
ie.
$$q_i(Y)\mapsto(\gamma_5\otimes1)q_i(Y);\;\;
\bar q_i(Y)\mapsto-\bar q_i(Y)(\gamma_5\otimes1);\;\;
\sigma({\tilde x})\mapsto-\sigma({\tilde x}).\eqno(2.9b)$$
It is this symmetry, corresponding to the continuum form
$\psi\mapsto\gamma_5\psi$, $\bar\psi\mapsto-\bar\psi\gamma_5$, which is
spontaneously broken at strong coupling, signalled by the appearance of a
non-vanishing condensate $<\bar\chi\chi>$
or equivalently $<\bar q(1_4\otimes1_2)q>$. We
shall see in the next section that to leading order in $1/N_f$ the lattice
formulation (2.1) gives predictions in agreement with those of the continuum.

The action (2.1) was numerically simulated using the hybrid Monte Carlo
algorithm [16], in which the Grassmann fields are replaced by real bosonic
pseudofermion fields $\phi(x)$ governed by the action
$$S=\sum_{x,y}\sum_{i,j=1}^{N_f/2}{1\over2}\phi_i(x)\big(M^tM\big)_{xyij}^{-1}
\phi_j(y) + {N_f\over{4g^2}}\sum_{\tilde x}\sigma^2({\tilde x}),\eqno(2.10)$$
where
$$M_{xyij}={\cal M}_{xy}\delta_{ij}+\delta_{xy}\delta_{ij}{1\over8}
\sum_{<{\tilde x},x>}\sigma({\tilde x}).\eqno(2.11)$$
Note $M$ is strictly real. Integration over $\phi$ yields the
functional measure $\sqrt{{\rm det}(M^tM)}\equiv{\rm det}M$ if the determinant
of $M$ is positive semi-definite. This condition is fulfilled if $N_f/2$ is an
even number, even for $\mu\not=0$. The problem of complex determinants
associated with simulating gauge theories at finite density do not appear.
Further details of our simulation procedure are given in [5].

As well as measuring the expectation value of the scalar field $<\sigma>$ in
the simulation, which for our purposes is the order parameter of the
transition, we also monitored the chiral condensate $<\bar\psi\psi>$, the
energy density $<\epsilon>$, and the fermion number density $<n>$, which are
defined by
$$\eqalign{
-<\bar\psi\psi>&={1\over V}{\rm tr}S_F ={1\over V}<{\rm tr}M^{-1}>,\cr
<\epsilon>=-{1\over V_s}{{\partial\ln Z}\over{\partial\beta}}
&={1\over V}{\rm tr}\partial_0\gamma_0S_F ={1\over2V}
<\sum_xe^\mu M^{-1}_{x,x+\hat0}-e^{-\mu}M^{-1}_{x,x-\hat0}>,\cr
<n>=-{1\over{V_s\beta}}{{\partial\ln Z}\over{\partial\mu}}
&={1\over V}{\rm tr}\gamma_0S_F ={1\over2V}
<\sum_xe^\mu M^{-1}_{x,x+\hat0}+e^{-\mu}M^{-1}_{x,x-\hat0}>.\cr}\eqno(2.12)$$
Here $V_s$ is the spatial volume, $\beta$ the inverse temperature, and
$V=V_s\beta$ the overall volume of spacetime. The final expression in each case
is the quantity measured in the simulation, using a noisy estimator to
calculate the matrix inverses.
\vskip1truecm

\noindent{\bf3. Mean Field Analysis}

In this section we calculate the order parameter $\Sigma=<\sigma>$ as a
function of coupling $g$, chemical potential $\mu$, and temperature $T$
($\equiv1/\beta$), to leading order in $1/N_f$. Since the limit $N_f\to\infty$
suppresses fluctuations around the saddle point solution, this is equivalent
to a mean field treatment. We will work       both in the continuum and using
the specific lattice regularisation (2.1). If diagrams of $O(1/N_f)$ and
beyond are ignored, the only contribution to $\Sigma$ comes from a simple
fermion loop tadpole, and we determine $\Sigma$ self-consistently using the
gap equation:
$$\Sigma=-g^2<\bar\psi\psi>={g^2\over V}{\rm tr}S_F(\mu,T,    \Sigma),
\eqno(3.1)$$
where in the Euclidean formulation the fermi propagator $S_F$ is given by
$$S_F^{-1}(k;\mu,T,\Sigma)=i\gamma_0(k_0-i\mu)+\sum_{\nu=1,2}ik_\nu\gamma_\nu
+\Sigma.\eqno(3.2)$$
For non-zero temperatures the alllowed values of $k_0$ are quantised as
$$k_0=(2n-1)\pi T,\;\;\;\;n\in{\rm Z},\eqno(3.3)$$
ie. with antiperiodic boundary conditions in the finite temporal direction.
Collecting together equations (3.1-3)   we arrive at
$${1\over g^2}=4T\sum_{n=-\infty}^\infty\int{{d^2p}\over{(2\pi)^2}}
{1\over{\left((2n-1)\pi T-i\mu\right)^2+p^2+\Sigma^2}}.\eqno(3.4)$$

The manipulations from here are standard [10,12]. First one resums
over $n$ using the Poisson formula:
$${1\over g^2}=4T\sum_{m=-\infty}^\infty\int{{d^2p}\over{(2\pi)^2}}
\int_{-\infty}^\infty d\phi{{e^{2\pi im\phi}}\over{[(2\phi-1)\pi T-i\mu+iE]
[(2\phi-1)\pi T-i\mu-iE]}}.\eqno(3.5)$$
Here $E=\sqrt{p^2+\Sigma^2}$. If $\mu<\Sigma$, then the poles in the integrand
lie on opposite sides of the integration contour for all values of $p$, and
the integral over $\phi$ is easily performed to yield
$${1\over g^2}=\int_\Sigma^\infty{{dE}\over\pi}\left[1-
{1\over{e^{\beta(E-\mu)}+1}}-{1\over{e^{\beta(E+\mu)}+1}}\right].\eqno(3.6)$$
If, however, $\mu>\Sigma$ we must take care, since for certain values of $p$,
both poles in the integrand of (3.5) will lie to the same side of the contour.
We find in this case
$${1\over g^2}=\int_\mu^\infty{{dE}\over\pi}\left[1-{1\over{e^{\beta(E-\mu)}+1}
}\right]+\int_\Sigma^\mu{{dE}\over\pi}{1\over{e^{\beta(\mu-E)}+1}}-
\int_\Sigma^\infty{{dE}\over\pi}{1\over{e^{\beta(\mu+E)}+1}}.\eqno(3.7)$$
We now eliminate $g$ in favour of $\Sigma_0$, the order parameter at zero
temperature and chemical potential, using equation (3.6) at $T=\mu=0$. This
gives an implicit equation for $\Sigma$ in terms of a physical scale
$\Sigma_0$, with no reference to any UV cutoff. For $\mu<\Sigma$ we obtain
$$\Sigma_0-\Sigma=T\left(\ln(1+e^{-\beta(\Sigma-\mu)})
                        +\ln(1+e^{-\beta(\Sigma+\mu)})\right).\eqno(3.8a)$$
while for $\mu>\Sigma$:
$$\Sigma_0-\mu=T\left(\ln(1+e^{-\beta(\mu-\Sigma)})
                     +\ln(1+e^{-\beta(\mu+\Sigma)})\right).\eqno(3.8b)$$
In fact, these two equations are identical solutions for $\Sigma(\mu,T)$,
and also demonstrate that curves of $\Sigma(\mu)$ at constant $T$ are
symmetric under reflection in the line $\Sigma=\mu$. This result is
peculiar to three spacetime dimensions. Equation (3.8a) was first
derived in [12].

Equation  (3.8) gives          a complete solution for $\Sigma(\mu,T)$ in terms
of $\Sigma_0$. Since $\Sigma\to0_+$ smoothly, the symmetry-restoring transition
is second order throughout the $(\mu,T)$ plane, except for one isolated
point, as we shall see. To obtain the equation for the
critical line in this plane we set $\Sigma=0$ in (3.8) to get the curve:
$$1-{\mu\over\Sigma_0}=2{T\over\Sigma_0}\ln(1+e^{-\beta\mu}).\eqno(3.9)$$
At zero chemical potential, therefore, we predict a chiral symmetry-restoring
transition at a critical temperature
$$T_c={\Sigma_0\over{2\ln2}}\simeq0.72\Sigma_0.\eqno(3.10)$$
The gap equation in the broken phase in this limit is the $\mu=0$ limit of
(3.8):
$$\Sigma_0-\Sigma=2T\ln(1+e^{-\beta\Sigma}).\eqno(3.11)$$
At zero temperature we find $\Sigma=\Sigma_0$ independent of $\mu$ up to a
critical value
$$\mu_c=\Sigma_0,\eqno(3.12)$$
at which point there is a discontinuous drop to zero, ie. at this isolated
point the transition is first order. For small excursions into the $(\mu,T)$
plane we find from (3.8)
$${{\partial\Sigma}\over{\partial\mu}}\vert_{T\to0}=
\displaystyle{\lim_{T\to0}}-{{\sinh\beta\mu}\over{\sinh\beta\Sigma}}
\simeq-e^{\beta(\mu-\Sigma)},\eqno(3.13)$$
ie. the slope of the surface $\Sigma(\mu,T)$ diverges in an essentially
singular way as $\mu\to\mu_c$, $T\to0$. So, the mean field analysis
predicts a first order transition for $T=0$, which becomes second order
as soon as $T>0$.

Using a similar route we can also calculate the fermion number density
$<n>$ in the broken phase starting from (2.12). For $\mu<\Sigma$ we find
$$<n>={{\Sigma T}\over\pi}\ln{(1+e^{-\beta(\Sigma-\mu)})\over
                              (1+e^{-\beta(\Sigma+\mu)})} -
{{2T^2}\over\pi}\sum_{k=1}^\infty(-1)^k{{e^{-\beta k\Sigma}\sinh\beta k\mu}
                   \over k^2},\eqno(3.14)$$
whereas for $\mu>\Sigma$:
$$<n>={{\mu^2-\Sigma^2}\over{2\pi}}+{{\Sigma T}\over\pi}\ln
    {{(1+e^{-\beta(\mu-\Sigma)})}
\over{(1+e^{-\beta(\mu+\Sigma)})}}-{{2T^2}\over\pi}\sum_{k=1}^\infty
(-1)^k{{(1-e^{-\beta k\mu}\cosh\beta k\Sigma)}\over k^2}.\eqno(3.15)$$
In the symmetric phase we recover the usual expression for a two-dimensional
relativistic free fermi gas:
$$<n>={\mu^2\over{2\pi}}-{{2T^2}\over\pi}\sum_{k=1}^\infty(-1)^k
{{(1-e^{-\beta k\mu})}\over k^2}.\eqno(3.16)$$
In the limit $T\to0$ we see that fermion density is strongly suppressed for
$\mu<\Sigma$, then jumps discontinuously and continues to rise
quadratically with $\mu$ as soon as $\mu$ excceds
$\Sigma$. As required, $<n>$ vanishes for all $T$ when $\mu=0$.

We have also studied the gap equation using an explicit UV regularisation
defined by the lattice action (2.1). This is useful for comparison with the
numerical results, particularly so that we can determine whether any
differences with the continuum predictions arise from genuine $1/N_f$
corrections (ie. departure from mean field behaviour), or simply from the fact
that on a finite lattice it is impossible to attain the thermodynamic limit.
This latter point arises because lattice simulations at non-zero temperature
are generally accomplished using a system with $N_\tau$  lattice points in the
temporal direction, with $N_\tau \ll N$,   the spatial dimension. In the work
presented here $N_\tau$  ranges from 6 to 12, for $N  =36$. Clearly the main
effect is that the sum over Matsubara frequencies in the lattice gap equation
is truncated at a rather  small value of $n$.

Using the free fermion action in the form (2.6) we evaluate equation (3.1)
to yield the lattice gap equation on a system of infinite spatial extent:
$${1\over g^2}={8\over N_\tau} \int_{-\pi/2}^{\pi/2}{{d^2 k}\over{(2\pi)^2}}
\sum_n{1\over{{1\over2}\left\{1-\cos\left({{2\pi(2n-1)}\over N_\tau}\right)
\cosh 2\mu-i\sin\left({{2\pi(2n-1)}\over N_\tau}\right)\sinh 2\mu\right\}
+\sum_{\nu=1,2}\sin^2k_\nu+\Sigma^2}},\eqno(3.17)$$
where $\sum_n$ defines a sum running from $-{N_\tau \over4}+{1\over2}$ to
${N_\tau \over4}-{1\over2}$ if $N_\tau/2$        is odd or
$-{N_\tau \over4}+1$ to ${N_\tau \over4}$ if $N_\tau/2$        is even (note
$N_\tau$  must be even in order for staggered fermions to be defined). The
integral over $k$ can be done via Schwinger parameterisation to yield
$${1\over g^2}={2\over N_\tau} \int_0^\infty  d\alpha
e^{-\alpha({3\over2}+\Sigma^2)}I_0^2({\textstyle  {\alpha\over2}})
\sum_n\exp\left[{\alpha\over2}\cos\left({{2\pi(2n-1)}\over N_\tau}\right)
\cosh 2\mu\right]\cos\left[{\alpha\over2}
\sin\left({{2\pi(2n-1)}\over N_\tau}\right)\sinh 2\mu\right],\eqno(3.18)$$
which is now in a form suitable for numerical quadrature. $I_0$ is the modfied
Bessel function. Unfortunately the integral over $\alpha$ in (3.18) only
converges for $2\mu\leq \cosh^{-1}(\sec2\pi/N_\tau )$:
for values of $\mu$ greater
than this, although (3.17) is convergent, it is no longer possible to cast it
into convenient form.

Figure 3.1 shows a comparison of $\Sigma(\mu)$ calculated using the continuum
solution (3.8)   and the lattice solution (3.18) evaluated at inverse coupling
$1/g^2=0.75$ on a lattice with $N_\tau =6,8,10$ and 12. In practice we solve
(3.18) for $1/g^2$ as a function of $\Sigma$ and invert using interpolation.
The continuum solution sets $\beta=N_\tau$  and uses a value for $\Sigma_0$
given by the lattice gap equation at zero temperature and chemical
potential:
$${1\over g^2}=\int_0^\infty d\alpha e^{-\alpha({3\over2}+\Sigma_0^2)}
I_0^3({\textstyle  {\alpha\over2}}).\eqno(3.19)$$

We see that the agreement between the two is fair, with the lattice results
always lying below the continuum ones. As discussed above, we ascribe this
difference to the finite number of thermal modes available on the lattice.
\vskip1truecm

\noindent
{\bf 4.  Non-zero Chemical Potential Simulation Results}

We first studied the theory at non-zero chemical potential on a symmetric
lattice.  In ref. [5] we obtained accurate results at vanishing chemical
potential for lattices ranging in size from $8^3$ through $20^3$.  The vacuum
expectation value of $\sigma$,
and its susceptibility were measured over a range
of coupling extending from $1/g^2 \sim 0.5$
to 1.2.  Good agreement with large $N_f$
scaling laws were found for $1/g^2$ ranging from 0.70 to 1.1 and the critical
point in the infinite volume limit was estimated to be
$1/g_*^2 = 0.975-1.000$.  So, the scaling window of the lattice formulation
lying in the chirally asymmetric phase was seen to be $0.70  \leq 1/g^2
\leq 1.00$.  These results
 led us to simulate the model with $\mu\not=0$ on a
$20^3$ lattice at couplings $1/g^2
= 0.70, 0.75$ and 0.80.  Short exploratory runs indicated that larger lattices
would be needed to push $1/g^2$ closer to the critical point.  In Table 1 we
show the simulation results for various $\mu$ at $1/g^2 = 0.70$,
measurements of
$\Sigma= <\sigma>$,
the energy density $<\epsilon>$, and the induced ground state fermion number
density $<n>$ are recorded.  We also show the number of trajectories of the
hybrid Monte Carlo algorithm at each point.  The huge number of
trajectories relative to state-of-the-art lattice QCD simulations with
dynamical fermions was possible because of the three dimensional character
of the model and its relatively simple form, a random scalar field coupled
to a fermionic scalar density.  This last feature led to a conjugate
gradient routine which converged with an order of magnitude fewer sweeps
than typically needed
in lattice QCD simulations.  The results recorded in Table 1 are plotted in
Figure 2 where we see a jump discontinuity in $\Sigma$
vs. $\mu$ of 0.275 as $\mu$ varies
from 0.39375 to 0.3941.  Note from the table that we were particularly
careful to accumulate good statistics near the transition.  In Figure 3 we
show the induced fermion number $<n>$ plotted against $\mu$ which shows a clear
jump over the same coupling range.  We believe that the nonvanishing values
of $<n>$ recorded for
$\mu<0.3941$ are finite size (temperature) effects.  Finite size
effects will be
discussed further in Section 6 below.  The error bars in the table and
plotted in the figures come from standard binning procedures.  It was
possible to "confirm" these error estimates in many cases by running the
algorithm for an extra several thousand trajectories and reproducing
average values and variances.

As shown in Table 2 the simulation was repeated at $1/g^2 = 0.75$, slightly
closer to the bulk critical point.  The critical chemical potential shifted
to
$\mu_c = 0.32 - 0.3225$
and discontinuities were again observed in   $\Sigma$ vs. $\mu$ (Figure
4) and $<n>$ vs. $\mu$ (Figure 5).  The discontinuities were slightly more
difficult to measure because of the proximity to the critical point which
reduces the size of physical observables when measured in lattice units.

Runs were also completed at $1/g^2 = 0.80$ and a value of $\mu_c = 0.250(5)$
was measured.  Less extensive data was taken here as reflected in the larger
error bar.  We identified $\mu_c$ by plotting time histories of
$\Sigma$  ($\Sigma$ vs.
computer time) and noting that for $\mu<\mu_c$ maintained a non-zero value
while for $\mu>\mu_c$ evolved to zero and fluctuated around it.  These results
were less quantitative than the $1/g^2 = 0.70$ and 0.75 simulations because
critical slowing down was affecting the efficiency of the algorithm.  In
fact, we abandoned an attempt to study $1/g^2 = 0.85$ because of severe
critical slowing down and the possibility of large finite size effects
invalidating an estimate of $\mu_c$ on a $20^3$ lattice.

It is interesting to analyze the measurements of $\mu_c$
at $1/g^2 = 0.70, 0.75$ and
0.80 for two additional purposes.  The first is to estimate $\mu_c$ in physical
rather than lattice units.  The natural way to do this is to record the
ratio of $\mu_c$ to  $\Sigma_0$
measured at the same coupling as the finite-m transition.
At large $N_f$, $\Sigma_0$
is essentially the dynamical fermion mass and Mean Field
theory predicts that $\mu_c/\Sigma_0= 1.0$.
{}From ref.[5] we have the values of $\Sigma_0$
on a $20^3$ lattice ($\mu = 0$) at $1/g^2 = 0.70 (\Sigma_0 = 0.432)$,
$1/g^2 = 0.75 ( \Sigma_0 = 0.346)$
and $1/g^2 = 0.80 ( \Sigma_0 = 0.262)$.
The ratios $\mu_c/\Sigma_0$ at each coupling are plotted
in Figure 6.  The error bars come almost exclusively from the $\mu_c$
measurements -- the $20^3$ $\mu = 0$ measurements of  $\Sigma_0$
recorded in ref.[5] were
very accurate indeed.  The $1/g^2 = 0.80$ result for $\mu_c$ is particularly
uncertain.  Nonetheless, as the critical point is approached the curve
strongly suggests that $\mu_c/\Sigma_0$
approaches unity in accord with Mean Field
theory, although a departure downwards from this value as a result of $1/N_f$
corrections cannot be ruled out.

Our last use for this data is for a calculation of the critical index $\nu$,
the critical index of the correlation length.  Since the critical chemical
potential $\mu_c$ is a dimensionful parameter coupled to a conserved current in
the Lagrangian, it undergoes no renormalization due to $1/N_f$ corrections,
and should scale as a physical mass as the critical point is approached,
i.e. vanishing in lattice units with the exponent $\nu$

$$
\mu_c = C \bigl(1/g_*^2 - 1/g^2\bigr)^\nu
\eqno(4.1)
$$
{}From ref.[5] $1/g_*^2 = 0.975 - 1.00$.  In Figure 7 we show a
plot of $\ln\mu_c$ vs.
$\ln \bigl(1/g^2 - 1/g_*^2\bigr)$
with choice $1/g_*^2 = 0.975$.  A linear fit is good and it gives
$\nu= 1.00(5)$ to be compared with the first two terms of the large-$N_f$
expansion $(\nu= 1 + 8/3N_f\pi^2= 1.0225..$, for $N_f = 12)$.
A similar plot for
$1/g_*^2 = 1.00$ gives $\nu= 1.10(5)$.  So, in summary we have

$$
\nu=1.05(10)
\eqno(4.2)
$$
in excellent agreement with the large-$N_f$ analysis.  This analysis and Eq.
(4.2) could certainly be pursued more systematically with greater control,
but consistency between the analytic and numerical approaches to this
problem is our only goal here.

A systematic search for $1/N_f$ effects in our simulation results, or in more
accurate simulations on larger lattices that might be done in the future,
would require more calculations than done here.  For example, if we take
our results for $\mu_c/\Sigma_0$
at face value, ie. an increasing function of $1/g^2$ as
$1/g^2\to 1/g_*^2$, then we would predict the exponent $\nu$
to be {\it smaller} than the
exponent $\beta_m$
governing the scaling of  $\Sigma_0$, independently of any estimate of
the exact value of $1/g_*^2$.  This is at variance with the $O(1/N_f)$
corrections of eq.(1.2) which predict

$$
{ {\mu_c(g)}\over{\Sigma_0(g)} }\sim
{ {(1/g_*^2 - 1/g^2)^\nu}\over{(1/g_*^2 - 1/g^2)^{\beta_m}} }\sim
(1/g_*^2 - 1/g^2)^{8\over{3N_f\pi^2}}
\eqno(4.3)
$$
Numerically $8/3N_f\pi^2 = 0.0225$ for $N_f = 12$,
and since $1/g_*^2 - 1/g^2$ varies
from 0.3 to 0.2 over the region of couplings explored here, the right hand
side of eq.(4.3) varies by less than a percent.  Clearly much greater
precision is needed before we can interpret the trend of Figure 7 to $1/N_f$
effects, which would require large corrections at $O(1/N_f^2)$ and beyond.
\vskip1truecm

\noindent{\bf 5.  Non-Zero Chemical Potential and Temperature}

In ref.[5] we studied the four Fermi model at non-zero temperature by
simulating it on asymmetric lattices $N_\tau\times N^2$ with
$N\gg N_\tau$ and $N_\tau$ ranging from
4 to 12.  A second order chiral transition was discovered with a critical
temperature measured in physical units of $T_c/\Sigma_0 \approx 0.70(5)$.
The result
was in good agreement with Mean Field theory which predicts
$T_c/\Sigma_0= (2\ln 2)^{-1} =0 .721$.
This result and the zero temperature, non-zero
chemical potential result of Section 4, were then extended to map out the
$\mu-T$ phase diagram.  We simulated $6\times 36^2, 8\times 36^2,
10\times 36^2$ and $12\times 36^2$
lattices and obtained mc on each of these lattices by measuring $\Sigma$
vs. $\mu$.
Each lattice was simulated at a coupling $1/g^2 = 0.75$ because this coupling
lay in the scaling window of the lattice Lagrangian and because of the
success of Section 4.  The curves of $\Sigma$ vs. $\mu$
are shown in Figure 8.  The
statistics at each point are comparable with Table 1 and the error bars on
each point in the figure are smaller than the symbols themselves.  The
precise values of $\mu_c$ for each $N_\tau$
are recorded in Table 3 with error bars.
As usual, vacuum tunnelling (flipflops between  $\pm\Sigma$ values) limited our
precision on mc measurements.  It is interesting to compare Fig. 8 with its
Mean Field counterpart in Fig. 1.  The qualitative features of both plots
are identical, but the scale of $\Sigma$ and $\mu$
values coming from the computer
experiment are consistently below the Mean Field predictions.

The results in Table 3 can be converted to physical points on a
$\mu_c/\Sigma_0$ vs. $T_c/\Sigma_0$
phase diagram by recalling the value of  $\Sigma_0$ at $1/g^2 = 0.75$ at zero
temperature,  $\Sigma_0= 0.346$.  A bit of arithmetic and the fact that the
temperature is related to the temporal extent of the lattice
$N_\tau$, $T = N_\tau^{-1}$,
produces the phase diagram of Figure 9.  The solid line is the Mean Field
phase boundary between the low-$T$, low-$\mu$ symmetry broken phase and the
chiral symmetry restored phase.  Note that the point at $T_c/\Sigma_0 =0.482$
and $\mu_c/\Sigma_0 = 0.564$
comes from the smallest lattice $N_\tau = 6$ and is subject to the
largest finite size correction.  The data point at $m = 0$ comes from ref.[5].

It would be interesting to determine the order of the phase transition in
Figure 9 away from the $\mu =0$ and $T = 0$ boundaries to test the Mean Field
prediction of a line of second order transitions inside the plane.  Our
attempts to do this used time histories of $\Sigma$ to search for two-state
signals either visually or through histogramming.  The results were not
conclusive, however.  In light of finite size effects to be discussed
below, we believe that larger $N_\tau$ values would actually be needed to
accomplish this goal.
\vskip1truecm

\noindent{\bf 6. Finite Size and Finite Mass Effects}

Finally, we did two additional simulations motivated as much by lattice QCD
simulations as our interest in three dimensional four Fermi models.  It is
clear from the lattice Mean Field discussion in Section 3 above that even
in the large $N_f$ limit where fluctuations are suppressed relatively large
lattices are needed to simulate the theory's actual critical behavior.  We
show this effect in Figure 10 which shows the induced fermion number
plotted against $\mu$ for a $12^3$ lattice at $1/g^2 = 0.70$.  The
discontinuous
transition seen on a $20^3$ lattice in Figure 3 is replaced by a relatively
smooth curve.  A careful search for metastability in the region
$0.35<\mu <0.40$ revealed none.

It is also of interest to add a small explicit chiral symmetry breaking
fermion bare mass to the Lagrangian and record its tendency to smooth out
the transition.  In fact, even a very small bare fermion mass obscures the
first order transition of Figure 3 entirely.  This effect is shown in
Figure 11 where the theory is again simulated on a $20^3$ lattice at
$1/g^2 =0.70$
and the induced fermion charge is measured with the bare mass of the
dynamical fermion chosen to be either $m = 0.01$ or 0.005 in lattice units.  A
comparison with Figure 3 shows that $< n (\mu,m) >$
decreases as $m$ is increased
from zero; in either phase this reflects the fact that more energy is
always needed to excite fermion states from the vacuum once a bare mass is
introduced.  The average value of the sigma field is shown in Figure 12 for
the same conditions.  In neither case did we find any evidence for
metastability, although, as shown in the figures, a very fine grid of $\mu$
values were simulated.  One can also use the $m = 0.01$ and 0.005 results in
Figure 12 to linearly extrapolate the average values of
$\Sigma$ to zero fermion
mass at each value of $\mu$.  One can read off the figure that this procedure
predicts a critical point at $\mu_c=0.42 - 0.43$, considerably higher than the
actual value of 0.3933(8) shown in Figure 1.
A more sophisticated extrapolation method appears required to achieve
quantitative results.  QCD enthusiasts may find similar problems in their
more complicated systems.
\vskip1truecm

\noindent{\bf7. Conclusions}

In this study we have presented results of numerical simulations of an
interacting relativistic field theory near its continuum limit at non-vanishing
baryon-number density on far larger systems than have been possible hitherto
[3]. The main result of our work is that the data we have is
in excellent qualitative agreement with analytic predictions based on the
$1/N_f$ expansion. We observe that in the regime $1/g^2<1/g^2_*$ where chiral
symmetry is spontaneously broken at zero temperature and density, for
sufficiently large chemical potential $\mu$ there is a phase transition
which restores the symmetry. For $T=0$ the transition is first order: for
$T>0$ the transition appears continuous. Both results are
consistent with the leading order $1/N_f$ prediction that the transition
at $(\mu/\Sigma_0=1,T=0)$ is an isolated first order point; however, the
possibility of first order behavior persisting for values of $T>0$
cannot be excluded.
In all cases our measured
values of $\Sigma(\mu,T)$ fall slightly below the predictions of Sec. 2 --
it will require further systematic study to establish whether this is due to
finite volume effects or to genuine $1/N_f$ corrections.

It is interesting to contrast this situation with that of the two-dimensional
Gross-Neveu model, where the leading order $1/N_f$ expansion predicts a
very rich phase diagram with a tricritical point in the $(\mu,T)$ plane
separating first and second order symmetry  restoring transitions [17].
Once quantum fluctuations are switched on, however, (ie. $N_f<\infty$), the
symmetry restoration is dominated by condensation of kink -- anti-kink states,
which are not present in the $1/N_f$ expansion,
and the transition becomes first order everywhere apart from an isolated
second-order point at $\mu=0$ [11].

We can speculate on whether the first order nature of the $T=0$ transition
of the three-dimensional model remains stable as $T$ is increased from zero,
ie. whether there is in fact a tricritical point somewhere in the $(\mu,T)$
plane in this case, around which $\Sigma$ exhibits power-law scaling rather
than the essential singularity predicted in eqn. (3.13). Studies on much larger
systems, enabling us to probe sufficiently low temperatures  free of finite
volume effects, would be required to locate such a point unambiguously.
However, we note supporting evidence for the existence of such a point from
a comparison on Figs. 3 and 10, showing plots of $<n>$ vs. $\mu$ on $20^3$
and $12^3$ systems respectively. The major difference in the plots occurs in
the broken phase, where $<n>$ is considerably suppressed on the larger system,
supporting the assertion  made in Sec. 4 that the signal in this phase is a
finite volume effect (mean field theory predicts $<n>=0$ for $T=0$ in the
broken phase, eqn. (3.14)). Even on the $20^3$ lattice the presence of a
signal shows that the system is experiencing a small but non-zero effective
temperature: however the symmetry-restoring transition on this lattice is
clearly first order, suggesting that the discontinuous nature of the
transition is stable some way into the $(\mu,T)$ plane. Studies of the
$O(1/N_f)$ corrections would be of value here: a first order
transition would manifest itself in an extra non-trivial solution of the
gap equation for $\mu,T>0$, implying an extra extremum in the effective
potential for $\Sigma$ [17].

By studying the scaling of the critical chemical potential $\mu_c$ as a
function of coupling, we have been able to extract the correlation length
critical exponent $\nu$. This is unusual, because we have only measured
bulk properties of the system, and not any two-point correlations. Of course,
our analysis implicitly relies on the theory's renormalizability [5,6]. Our
result is in good agreement with the leading order $1/N_f$ prediction $\nu=1$,
and our precision is slightly better than that obtained through finite-size
scaling studies [5]. As always in these measurements, the major uncertainty is
the location of the bulk critical point $1/g^2_*$. Clearly from our data much
greater precision will be required to probe $O(1/N_f)$ corrections.

Finally, our studies on small volumes and with non-zero bare masses highlight
the difficulties that must be faced when trying to understand the critical
behavior of the chiral/thermodynamic limit by extrapolation from systems
away from these limits. In this case the first order nature of the zero
temperature transition is obscured -- this should be a warning for the lattice
QCD community (if one is still needed!) that a quantitative understanding of
the
behavior of QCD at non-vanishing density lies along a difficult road.
\vskip1truecm

\noindent{\bf Acknowledgments}

The calculations reported here used the resources of the National Energy
Research Supercomputer Center, the Pittsburgh Supecomputer Center and the
Materials Research Laboratory at the University of Illinois at
Urbana-Champaign under grant NSF DMR89-20538.  JBK is supported in part by
NSF-PHY87-01775.  SJH is supported in part by an SERC Advanced Fellowship,
and we would like to thank Bill Wyld for helpful discussions in the early
stages of this work.
\vfill\eject

\noindent{\bf References}

\noindent
[1] D.A. Kirzhnits and A.D. Linde, Ann. Phys. {\bf101} (1976) 195.

\noindent
[2] C. Bernard, T.A. DeGrand, C. DeTar, S. Gottlieb, A. Krasnitz, M.C. Ogilvie,
R.L. Sugar and D. Toussaint, Phys. Rev. Lett. {\bf68} (1992) 2125;\hfill\break
T.H. Hansson and I. Zahed, Nucl. Phys. {\bf B374} (1992) 277;\hfill\break
G.E. Brown, A.D. Jackson, H.A. Bethe and P.M. Pizzochero, SUNY preprint
NTG-92-06 (1992).

\noindent
[3] I.M. Barbour, talk presented at {\sl Lat '91\/}, Tsukuba, Japan (to
appear in Nucl. Phys. B(Proc. Suppl.)).

\noindent
[4] D.J. Gross and A. Neveu, Phys. Rev. {\bf D20} (1974) 3235.

\noindent
[5] S.J. Hands, A. Koci\'c and J.B. Kogut, CERN preprint.

\noindent
[6] S.J. Hands, A. Koci\'c and J.B. Kogut, Phys. Lett. {\bf B273} (1991) 111.

\noindent
[7] S.-K. Ma, {\sl Modern Theory of Critical Phenomena\/} (Benjamin, New York,
1976).

\noindent
[8] N.D. Mermin and H. Wagner, Phys. Rev. Lett. {\bf17} (1966)
1133;\hfill\break
S. Coleman, Commun. Math. Phys. {\bf31} (1973) 259.

\noindent
[9] T. Jolic\oe ur and A. Morel, Nucl. Phys. {\bf B262} (1985) 627;\hfill\break
J. Gasser and H. Leutwyler, Phys. Lett. {\bf188B} (1987) 477.

\noindent
[10] R.F. Dashen, S.-K. Ma and R. Rajamaran, Phys. Rev. {\bf D11} (1975) 1499.

\noindent
[11] F. Karsch, J.B. Kogut and H.W. Wyld, Nucl. Phys. {\bf B280 [FS18]} (1987)
289.

\noindent
[12] B. Rosenstein, B.J. Warr and S.H. Park, Phys. Rev. {\bf D39} (1989) 3088.

\noindent
[13] J.B. Kogut, H. Matsuoka, S.H. Shenker, J. Shigemitsu, D.K. Sinclair,
M. Stone and H.W. Wyld, Nucl. Phys. {\bf B225 [FS9]} (1983) 93;\hfill\break
P. Hasenfratz and F. Karsch, Phys. Lett. {\bf B125} (1983) 308.

\noindent
[14] Y. Cohen, S. Elitzur and E. Rabinovici, Nucl. Phys. {\bf B220} (1983) 102.

\noindent
[15] C.J. Burden and A.N. Burkitt, Europhys. Lett. {\bf3} (1987) 545.

\noindent
[16] S. Duane, A.D. Kennedy, B.J. Pendleton and D. Roweth, Phys. Lett.
{\bf B195} (1987) 216.

\noindent
[17] U. Wolff, Phys. Lett. {\bf157B} (1985) 303.
\vfill\eject

\noindent{\bf Table Captions}

\noindent
1. Chemical Potential $\mu$, ground state expectation value of the sigma
field $\Sigma$,
energy density $\epsilon$ and ground state expectation value of the   induced
fermion number $<n>$ on a $20^3$ lattice at coupling $1/g^2 = 0.70$.
The number of
trajectories for the hybrid Monte Carlo algorithm is    recorded in the last
column.

\noindent
2. Same as Table 1 except $1/g^2 = 0.75$.

\noindent
3. Measurement of $\Sigma$  vs. $\mu$ on $N_\tau\times N^2$
$(N = 36)$ lattices.  The coupling $1/g^2=0.75$.
\vfill\eject

\centerline{\bf Table 1}
\vskip5truemm
$$\vbox{\settabs\+\qquad.39375\qquad&\qquad.275(8)\qquad&\qquad.321(3)\qquad&
\qquad.017(2)\qquad&\quad Trajectories\quad&\cr
\+\hfill$\mu$\hfill&\hfill$\Sigma$\hfill&\hfill$\varepsilon$\hfill&\hfill
$<n>$\hfill&\quad Trajectories\quad&\cr
\bigskip
\+\hfill.10\hfill&\hfill.430(1)\hfill&\hfill.287(1)\hfill&\hfill.00013\hfill&
\hfill2500\hfill&\cr\smallskip
\+\hfill.20\hfill&\hfill.429(1)\hfill&\hfill.288(1)\hfill&\hfill.00015\hfill&
\hfill2500\hfill&\cr\smallskip
\+\hfill.30\hfill&\hfill.423(1)\hfill&\hfill.290(1)\hfill&\hfill.0013\hfill&
\hfill2500\hfill&\cr\smallskip
\+\hfill.35\hfill&\hfill.401(1)\hfill&\hfill.295(1)\hfill&\hfill.0036\hfill&
\hfill1500\hfill&\cr\smallskip
\+\hfill.36\hfill&\hfill.396(2)\hfill&\hfill.297(1)\hfill&\hfill.005(1)\hfill&
\hfill1500\hfill&\cr\smallskip
\+\hfill.37\hfill&\hfill.373(3)\hfill&\hfill.299(2)\hfill&\hfill.006(1)\hfill&
\hfill1500\hfill&\cr\smallskip
\+\hfill.38\hfill&\hfill.351(4)\hfill&\hfill.304(3)\hfill&\hfill.008(1)\hfill&
\hfill1500\hfill&\cr\smallskip
\+\hfill.39\hfill&\hfill.318(6)\hfill&\hfill.313(3)\hfill&\hfill.013(1)\hfill&
\hfill3000\hfill&\cr\smallskip
\+\hfill.3925\hfill&\hfill.307(8)\hfill&\hfill.316(3)\hfill&\hfill.015(2)\hfill&
\hfill3000\hfill&\cr\smallskip
\+\qquad.39375\qquad&\qquad.275(8)\qquad&\qquad.321(3)\qquad&\qquad.017(2)
\qquad&\hfill6000\hfill&\cr\smallskip
\+\hfill.3941\hfill&\hfill.00\hfill&\hfill.334(3)\hfill&\hfill.026(2)\hfill&
\hfill6000\hfill&\cr\smallskip
\+\hfill.395\hfill&\hfill.00\hfill&\hfill.335(2)\hfill&\hfill.026(2)\hfill&
\hfill3000\hfill&\cr\smallskip
\+\hfill.40\hfill&\hfill.00\hfill&\hfill.342(2)\hfill&\hfill.029(2)\hfill&
\hfill1500\hfill&\cr}$$
\vfill\eject

\centerline{\bf Table 2}
\vskip5truemm
$$\vbox{\settabs\+\qquad.3225\qquad&\qquad.346(1)\qquad&\qquad.300(1)\qquad&
\qquad.00035(70)\qquad&\quad Trajectories\quad&\cr
\+\hfill$\mu$\hfill&\hfill$\Sigma$\hfill&\hfill$\varepsilon$\hfill&\hfill
$<n>$\hfill&\quad Trajectories\quad&\cr
\bigskip
\+\hfill.10\hfill&\qquad.346(1)\qquad&\qquad.300(1)\qquad&\qquad.00035(70)
\qquad&\hfill2500\hfill&\cr\smallskip
\+\hfill.20\hfill&\hfill.340(2)\hfill&\hfill.302(1)\hfill&\hfill.00032(65)
\hfill&\hfill1500\hfill&\cr\smallskip
\+\hfill.30\hfill&\hfill.282(3)\hfill&\hfill.312(2)\hfill&\hfill.006(2)\hfill&
\hfill1500\hfill&\cr\smallskip
\+\hfill.31\hfill&\hfill.261(7)\hfill&\hfill.316(2)\hfill&\hfill.009(1)\hfill&
\hfill1500\hfill&\cr\smallskip
\+\hfill.32\hfill&\hfill.221(9)\hfill&\hfill.322(2)\hfill&\hfill.012(1)\hfill&
\hfill3000\hfill&\cr\smallskip
\+\qquad.3225\qquad&\hfill.00\hfill&\hfill.331(3)\hfill&\hfill.016(1)\hfill&
\hfill3000\hfill&\cr\smallskip
\+\hfill.325\hfill&\hfill.00\hfill&\hfill.330(3)\hfill&\hfill.015(1)\hfill&
\hfill3000\hfill&\cr\smallskip
\+\hfill.33\hfill&\hfill.00\hfill&\hfill.333(3)\hfill&\hfill.017(1)\hfill&
\hfill1500\hfill&\cr\smallskip
\+\hfill.335\hfill&\hfill.00\hfill&\hfill.336(3)\hfill&\hfill.020(1)\hfill&
\hfill3000\hfill&\cr\smallskip
\+\hfill.34\hfill&\hfill.00\hfill&\hfill.337(1)\hfill&\hfill.021(1)\hfill&
\hfill3000\hfill&\cr}$$
\vfill\eject

\centerline{\bf Table 3}
\vskip5truemm
$$\vbox{\settabs\+.19\qquad&.050(15)\quad&\quad.27\qquad&.063(20)\quad&\quad.285
\quad&.100(10)\quad&\quad.31\qquad&.030(15)\quad&\cr
\+\qquad\qquad$N_\tau=6$&&\qquad\qquad$N_\tau=8$&&\qquad\qquad$N_\tau=10$&&
\qquad\qquad$N_\tau=12$&&\cr\bigskip
\+\hfill$\mu$\hfill&\hfill$\Sigma$\hfill&\hfill$\mu$\hfill&\hfill$\Sigma$\hfill&
\hfill$\mu$\hfill&\hfill$\Sigma$\hfill&\hfill$\mu$\hfill&\hfill$\Sigma$\hfill&
\cr\bigskip
\+0\hfill&.259(2)\hfill&\quad0\hfill&.318(3)\hfill
&\quad0\hfill&.335(2)\hfill&\quad0\hfill&.342(2)\hfill&\cr\smallskip
\+.10\hfill&.236(3)\hfill&\quad.10\hfill&.310(3)\hfill
&\quad.20\hfill&.307(2)\hfill&\quad.10\hfill&.338(2)\hfill&\cr\smallskip
\+.15\hfill&.181(3)\hfill&\quad.20\hfill&.260(4)\hfill
&\quad.25\hfill&.258(3)\hfill&\quad.20\hfill&.322(2)\hfill&\cr\smallskip
\+.16\hfill&.167(3)\hfill&\quad.25\hfill&.175(4)\hfill
&\quad.26\hfill&.243(3)\hfill&\quad.25\hfill&.290(3)\hfill&\cr\smallskip
\+.17\hfill&.140(3)\hfill&\quad.26\hfill&.126(4)\hfill
&\quad.27\hfill&.202(3)\hfill&\quad.26\hfill&.281(3)\hfill&\cr\smallskip
\+.18\hfill&.090(8)\hfill&\quad.27\qquad&.063(20)\quad
&\quad.28\hfill&.192(3)\hfill&\quad.28\hfill&.241(3)\hfill&\cr\smallskip
\+.19\qquad&.050(15)\quad&\quad.28\hfill&.00\hfill
&\quad.285\quad&.122(4)\hfill&\quad.30\hfill&.138(7)\hfill&\cr\smallskip
\+.20\hfill&.00\hfill&&
&\quad.29\hfill&.100(10)\quad&\quad.31\qquad&.030(15)\quad&\cr\smallskip
\+&&&
&\quad.30\hfill&.00\hfill&\quad.32\hfill&.00\hfill&\cr}$$
\vfill\eject

\noindent{\bf Figure Captions}
\vskip5truemm

\noindent
1. Mean Field predictions for $\Sigma$ vs. $\mu$
curves for $1/g^2 = 0.75$.  Solid
lines   are solutions for the continuum Eq.(3.8) for $N_\tau = 6, 8, 10$ and 12
reading from the bottom left to the top right.  The dotted lines are
solutions of the discrete Eq.(3.18) (where convergent) for $N_\tau = 6, 8, 10$
and 12.  The difference between the two sets of curves gives a  measure
of discretization effects.

\noindent
2. Chiral order parameters   $\Sigma$ vs. chemical potential $\mu$
on a $20^3$ lattice at coupling $1/g^2 = 0.70$.

\noindent
3. Induced ground state fermion number $<n>$ vs. $\mu$  on a $20^3$ lattice at
$1/g^2 = 0.70$.  The dashed line is the mean field prediction.

\noindent
4. Chiral order parameters   $\Sigma$
vs. chemical potential $\mu$  on a $20^3$ lattice at coupling $1/g^2 =0.75$.

\noindent
5. Induced ground state fermion number $<n>$ vs. $\mu$  on a $20^3$
lattice at $1/g^2 = 0.75$.

\noindent
6. $\mu_c/\Sigma_0$ vs. $1/g^2$ for a $20^3$ lattice.

\noindent
7. $\ln\mu_c$ vs. $\ln (1/g_*^2 - 1/g^2)$ for $1/g^2 =0.70,0.75$ and 0.80
on a $20^3$ lattice with the bulk transition $1/g_*^2 = 0.975$.

\noindent
8. $\Sigma$ vs. $\mu$ for asymmetric lattices $N_\tau\times N^2$
with $N_\tau = 6, 8, 10$ and 12 and $N = 36$ at coupling $1/g^2 = 0.75$.

\noindent
9. The phase diagram, $\mu_c/\Sigma_0$ vs. $T_c/\Sigma_0$,
for the three dimensional Four-Fermi model.
The solid Mean Field line separates the chirally broken
phase at low $\mu$ and $T$ from the symmetric phase at large $\mu$ and $T$.

\noindent
10. The induced fermion number $<n>$ vs. $\mu$ at coupling $1/g^2 =0.70$
on a "small" $12^3$ lattice.

\noindent
11. $<n>$ vs. $\mu$ at $1/g^2 =0.70$ on a $20^3$ lattice for the
theory with a bare fermion mass term, $m = 0.01$ and $0.005$ in lattice units.

\noindent
12. $\Sigma$ vs. $\mu$ for the same parameters as in Fig. 11.

\end